# A Matter of Mindset? Features and Processes of Newsroom-based Corporate Communication in Times of Artificial Intelligence

Tobias Rohrbach & Mykola Makhortykh

Institute of Communication and Media Studies, University of Bern

*Last update: 9 July 2024*

Many companies adopt the corporate newsroom model to streamline their corporate communication. This article addresses why and how corporate newsrooms transform corporate communication following the rise of artificial intelligence (AI) systems. It draws on original data from 13 semi-structured interviews with executive communication experts in large Swiss companies which use corporate newsrooms. Interviews show that corporate newsrooms serve as an organisational (rather than spatial) coordination body for topic-oriented and agile corporate communication. To enable their functionality, it is crucial to find the right balance between optimising and stabilising communication structures. Newsrooms actively adopt AI both to facilitate routine tasks and enable more innovative applications, such as living data archives and channel translations. Interviews also highlight an urgent need for AI regulation for corporate communication. The article's findings provide important insights into the practical challenges and coping strategies for establishing and managing corporate newsrooms and how newsrooms can be transformed by AI.

**Introduction**

The digitalisation processes transform the corporate communication landscape by increasing the volume and variety of topics, channels, and stakeholders that business companies must manage (Vogler and Badham, 2023; Cornelissen, 2020; Plesner and Raviola, 2016). These challenges are amplified by the constant development of communication technologies, which creates new possibilities and risks for corporate communication practices (Brockhaus *et al*., 2022; Coombs and Holladay, 2023; Zerfass *et al*., 2020). The recent advances in artificial intelligence (AI) contribute to the "fourth industrial revolution" that may fundamentally change (corporate) communication strategies (Xu *et a*l., 2018).To adapt to new possibilities and risks, companies shift from the traditional corporate communication models, which are often characterised by rigidness and lack of coordination, to new organisational structures (Zerfass and Schramm, 2014; Brockhaus and Zerfass, 2022; Johansen and Esmann Andersen, 2012).

In this article, we examine how one of such novel communication structures—the corporate newsroom model—is adopted in the Swiss corporate environment and how it tackles the latest AI-driven communication innovations. Inspired by classic journalistic newsrooms, corporate newsrooms provide a central strategic and operative hub for organising communication regarding topics relevant to the company. Moss (2021, p. 27) defines corporate newsrooms as a "spatially combined unit for corporate communications [...] with separate responsibilities for topics and channels". By facilitating organised exchanges between stakeholders within and across departments on a regular basis (e.g. daily, weekly, or monthly), newsrooms obviate common problems of modern communication departments, such as disconnected intra-company silos (Bento *et al*., 2023; Hoffjann, 2024) and a lack of agility (Ragas and Ragas, 2021; Seiffert-Brockmann *et al*., 2021).

In theory, this integrated coordination allows companies to avoid an overload of information and communication channels and respond faster to the changing corporate environment (Johansen and Esmann Andersen, 2012). In practice, however, newsrooms do not exist as an ideal type and must be tailored to the specific national and corporate context. In the case of the Germany-Austria-Switzerland triangle, there is initial evidence of the active adoption of the newsroom model for corporate communication (Seiffert-Brockmann *et al.*, 2021; Moss *et al.*, 2019; Keel and Niederhäuser, 2016), albeit the exact implementations can vary substantially among individual companies. With the exception of a single handbook (Moss, 2021), there are currently no empirical assessments on how practitioners navigate key challenges involved in the implementation and management of corporate newsrooms; similarly, insights into the use of AI in corporate newsrooms are missing.

The aim of this study is twofold: to empirically examine the key structures and processes of corporate newsrooms in Switzerland and to identify applications and risks of AI for newsroom-based communication. To do so, we draw on original data from 13 semi-structured interviews with executive communication experts in large Swiss companies who have adopted a newsroom model. The results indicate that corporate newsrooms are understood as organisational (rather than spatial) coordination bodies for topic-oriented and agile corporate communication. The active adoption of AI within newsrooms (e.g. for living data archives and channel translations) prompts an urgent need for AI regulation in the context of corporate communication. By providing a detailed (and largely missing) insight into the practical challenges and coping strategies for establishing and managing corporate newsrooms, we contribute to bridging the gap between the growing use of corporate newsrooms and a lack of scholarship on their successful implementation and management.

**Literature review**

The digital transformation in corporate communication and related possibilities and risks have been the subject of a growing body of research in recent years. In contrast to the 20th century, when corporate communication relied on the journalistic media as a crucial communication intermediary, today's companies directly engage with their audiences through digital channels (Seiffert-Brockmann *et al*., 2021; Vogler and Badham, 2023). These new communication channels vary from social media platforms (Vernuccio, 2014; Capriotti *et al.*, 2021) to search engines (Dou *et al*., 2010) to mobile devices (Martinelli, 2019). Besides the expanded reach of corporate communication, digital channels also enable new formats of storytelling, ranging from the co-creation of brand stories via social media (Singh and Sonnenburg, 2012) to the exploitation of anthropomorphism via chatbot advertisements (Sun *et al*., 2024) and the use of virtual reality for marketing purposes (de Regt *et al*., 2021)

Under these circumstances, research is increasingly looking at new corporate communication models which account for and productively engage with the diversity of digital communication channels and modes of business storytelling (Plesner and Raviola, 2016; Zerfass and Link, 2023). One of these models is the corporate newsroom. Originally, the model was referred to as a social media newsroom and was introduced in 2007 (Zerfass and Schramm, 2014). It gained more traction in the 2010s with several studies (e.g. Holzinger and Sturmer, 2012; Zerfass and Schramm, 2014) discussing the potential of the model to adapt corporate communication to digital challenges. The core advantage of the model was its focus on aggregating a company's (social media) online content and thus serving as a single access point.

The advancement of digital corporate communication beyond social media platforms prompted the need to expand the social media newsroom model to the more general corporate newsroom. As elaborated by Moss (2021), the corporate newsroom goes beyond the aggregation of social media content; instead, it proposes a cross-platform and cross-departmental model of organising corporate communication. It often involves "the spatial restructuring of the communication departments" (Behrens *et al*., 2021, p. 16). The goal is to reassemble formerly dislocated groups of communicators from media relations, internal communication, advertisement, and sometimes marketing and bring them together, usually in an open-plan office.

Moss (2021) argues that the appeal of the corporate newsroom model is predicated upon three key advantages. First, the model increases communication effectiveness by making it more organised, avoiding task duplication and decreasing reaction time. Second, it makes internal communication more transparent and facilitates coordination across company departments. Third, it contributes to the company's outreach, as topics and channels are strategically streamlined to create a coherent representation of the company among internal and external stakeholders.

The process of adopting a newsroom model and the degree to which it allows realising the above-mentioned benefits, however, remains under-studied. So far, there are relatively few empirical studies on how corporate newsrooms are implemented. Most research has focused on the German context, from which the corporate newsroom model seems to have emerged. For example, Moss (2021) investigated the adoption of a newsroom by a large insurance company in Germany. Seiffert-Brockmann *et al*. (2021) examined the use of newsrooms in a selection of German and Austrian companies, whereas Buggisch (2021) analysed the newsroom model in DATEV, a large Germany-based software company. An exception from this Germany-centric approach is the study by Al-Warith and Moss (2021) on a corporate newsroom in a Dutch law enforcement institution. Though rich in insights, these case studies provide a limited understanding of processes and structures involved in implementing a corporate newsroom model on a more general level.

In contrast, a few studies have used surveys to scrutinise specific newsroom features. In a sample of 172 communication experts from German-speaking countries, Moss *et al*. (2019) found a substantial variety in the lived experiences of corporate newsrooms. While roughly half of experts (53%) report working or planning to work with newsrooms, their definitions of a newsroom vary. This ambiguity in the understanding is best illustrated by a study by Keel and Niederhäuser (2016), who surveyed 77 Swiss companies for the presence of five theoretically derived features of newsrooms: (1) structural separation of topic and channels, (2) a centralised open-plan office, (3) topic-based responsibilities, (4) daily meetings for coordination, and (5), a collectively shared topic management. While their results show a strong trend towards newsroom-based modes, many companies with newsrooms only implemented one or two of these criteria, whereas several companies without newsroom-based structures fulfil all the criteria. Zerfass and Schramm (2014) analysed the corporate communication departments of the 200 largest companies in Germany, the United States, and the United Kingdom and found limited presence of newsroom-based features (for social media communication), concluding that "many benefits ascribed to Social Media Newsrooms are not put into practice" (p. 89). In light of the scarce and mixed findings of, we propose the following more general exploration of the key features of newsroom-based corporate communication:

> ***RQ1***: *What structures and processes characterise corporate newsrooms in large Swiss companies?*

As discussed above, the corporate newsroom is a hybrid form of corporate-journalistic thinking used to respond to a changing technological communication landscape (Madsen and Andersen, 2023). Two distinct bodies of research inform how technological innovations shape newsroom-based communication. The first body has more explicitly focused on corporate communications and traced how new technologies amplified the volume of topics and channels of corporate communication (Vogler and Bedham, 2023; Zerfass and Schramm, 2014). In this context, AI has been linked to multiple benefits for corporate communication, ranging from increased discoverability through search engines (Nyagadza, 2022) and improved collection and processing of communication-related impact data (often referred to as social listening; Yavuz and Tire, 2023; Westermann and Forthmann, 2021) to prediction of communication crises (Coombs and Holladay, 2023) and optimisation of communication-related content production (e.g. advertisements; Spanos, 2021).

Such research has also uncovered risks of AI, such as new publicly visible crisis threats (also known as paracrises; Coombs and Holladay, 2012). One potential threat is AI being used to defame the company via AI-manipulated content (e.g. deepfakes; Westerlund, 2019). Another threat related to the malperformance of corporate AI systems due to negligence and the lack of quality control; such malperformance can result in outcomes viewed as offensive by certain stakeholders. Yet this body of scholarship does not address the impact of AI for newsroom-based communication specifically.

Though not focusing on corporate contexts, a second body of work provides insights into the integration of AI in journalistic newsrooms. Two observations seem particularly relevant to the corporate newsroom case. First, there is a high within- and between-newsroom variability in the use of AI for journalistic workflows, explained by varying technological capabilities and affinities. Expert interviews with journalists and editors-in-chief showcase mostly trial-and-error-based experimentation with AI technologies by highly tech-savvy individuals rather than a coordinated push towards innovation (Bastian *et al.*, 2021; Dralega, 2023; Munoriyarwa *et al.*, 2023; Reyes-de-Cózar *et al.*, 2022).

Second, Diakopoulos (2019) distinguishes three types of newsroom tasks with "different susceptibilities to automation" (p. 14). The skills-based tasks deal with more isolated work routines and involve clearly demarcated operations; these tasks can reliably be automated. Rules-based tasks require a more complex form of cognition and interpretation, such as text processing and generation, but the rise of large language models (LLMs) is now catalysing the automation of these tasks as well. Finally, knowledge-based tasks resist automation, at least for now, because they require background knowledge from which journalists draw contextualisation and sense-making. AI's role in journalistic newsrooms therefore depends on the mix of tasks in a given workflow. To what extent these tasks' composition in corporate newsrooms differs from journalistic contexts remains unclear. In light of the lack of research on how corporate newsrooms handle the opportunities and risks of AI, we thus pose the following second research question:

**RQ2**: *How do corporate newsrooms tackle the advancement of AI and how does it affect newsroom processes?*

**Methodology**

*Data collection*

To answer the research questions, we conducted 13 semi-structured interviews with experts responsible for managing the newsroom workflow. 10 interviewees work in the corporate newsrooms at large Swiss companies. We do not report the names of companies to ensure the anonymity of interviewees, but the companies are related to different business sectors, including public transport, insurance, energy, sport, public health, and public administration. The remaining three interviewees work in traditional journalistic newsrooms of Swiss print and broadcasting media organisations. The establishment of the contacts with the experts was facilitated by the Newsroom Communication AG, a consultancy specialising in implementing the newsroom model for the Swiss market.

Slightly more than half of interviewees (7) were men, which reflects the gender skewness of the leading managerial positions in the Swiss corporate environment. Most interviewees were in their 40s with one younger (early 30s) and a few older participants (50s-60s). Eight interviewees had extensive (10+ years) experience working in the journalistic sector before shifting to the corporate environment.

The interviews were conducted between April and May 2024 via video calls in Swiss German (i.e. the interviewees' native language). The average length of the interviews was 44 minutes; the shortest interview was 40 minutes, and the longest was one hour. The interview guide contained three blocks of open-ended questions regarding (1) experts' perception and understanding of the corporate newsrooms, (2) attitudes and use cases of AI, and (3) the benefits and risks of AI as well as its regulation. The interview guide, including the complete list of questions, is provided in the online appendix.

*Data coding and analysis*

All interviews were audio recorded and then automatically transcribed using a locally installed Whisper model, a state-of-the-art speech-to-text transcription tool from OpenAI (Radford *et al.*, 2023). Even though the Whisper model was not trained on Swiss German speech data, it shows satisfactory performance and usually requires a few manual revisions (Dolev *et al.*, 2024). The qualitative content analysis of the transcripts consisted of three steps. First, the material was structured by matching text passages to deductive categories (i.e. the question blocks of the interview guide). Second, the observations were summarised across interviews to eliminate redundancies and reduce complexity. Third, the observations were condensed inductively by comparing and contrasting observations within and between interviews.

**Results**

*Hybrid and dynamic understanding of corporate newsrooms*

The corporate newsroom does not exist in a singular sense. Echoing earlier studies, the interviews illustrate that, at least in the Swiss case, this concept carries diverse meanings. Nevertheless, interviewees' perceptions of what constitutes a corporate newsroom share three main features. The first feature is the *hybrid spatial realisation* of a newsroom. Around two thirds of the interviewees note that their companies see the newsroom as a physical location. The spatial realisation ranges from a simple meeting room (e.g. an office for daily meetings or a workplace for newsroom management) to a more sophisticated corporate environment with a dedicated infrastructure (e.g. a newsdesk and a screen). However, while most interviewees value the physical anchoring of their newsroom, they all stress that corporate newsrooms are, at their core, a hybrid or exclusively virtual communication model. This is apparent in the newsroom definition from one of the experts:

> For us, the corporate newsroom is, first and foremost, a collective attitude, a mindset, and an organisational structure for how we work together in communication matters.

In some cases, limited resources restrict the spatial realisation of the newsroom. Yet, the absence of physical newsroom structures is viewed not only as a disadvantage but also as an opportunity to focus on what really matters: the organisational dimension—that is, the mindset. This subordinate role of the spatial dimension, which is contrary to the ideal type of newsroom implementation, also has pragmatic reasons that are directly related to the Swiss-specific context. Due to the country's multilingualism, multiple local offices, and increasingly hybrid working conditions, focusing on physical newsrooms would complicate rather than facilitate communication coordination.

A second definitional feature is conceptual core newsroom principles that inform the organisational realisation of corporate newsrooms in specific contexts rather than relying on universally valid templates. Experts see their newsroom structures as reflecting the "singular character and context" of their companies. Accordingly, these structures vary substantially between specific companies and sectors and depend on the maturity of the newsroom (see more below). The only structural element that appears in all the interviews is the organisational structuring of newsroom activities according to topics and channels. Besides, the interviewees mentioned three conceptual principles that underlie the newsroom model and which, according to one expert, guide its "overall communicative effect":

- *Topic orientation*: Following the logic of "story first", the interviewees describe sharpening and transmitting the company's main expertise as the core communicative objective of the newsrooms. The transformation of corporate topics into stories marks the point of departure for newsroom-based communication. In this sense, topics are more than content clusters; they are the nexus of newsroom workflows that define how and for what purposes the newsroom staff engages with other company members and departments. According to several experts, a key advantage of centralising communication around topics is that it ensures coherent and effective communication.

- *Centralised control*: The multiplication of communication topics and channels raises risks of losing efficiency, for instance, when communication processes are duplicated by different departments. To enable better coordination, the newsrooms "break down silos" and "dismantle the individual backyards". Or, in the words of an interviewee: "Before, each subunit had its own agenda, its own routines. We have now consolidated all of that." Regular meetings serve as management tools, which allow all stakeholders to be involved and jointly decide on both short and long-term communication strategies.

- *Agility*: All the interviewees noted that their newsrooms rely on a designation of specific working roles–rather than rigid functions–to react effectively to unexpected developments (e.g. scandals, political decisions) and adapt to changing communication landscapes (e.g. crises, technological innovations). Newsroom roles presume that essential communication processes are clearly defined areas of responsibility to ensure the functioning of everyday communication. In addition to traditional roles such as topic or channel managers, the agility of corporate newsrooms is reflected in the establishment of needs-based roles that employees switch to when necessary. Some interviewees highlighted roles for bridging newsrooms to other departments or for ad hoc ("on the fly") execution of urgent communication interventions.

The adoption of these core ideas is based not on one-time managerial interventions but emerges as an ongoing practice. As a third definitional feature, we observe a consistent pattern across all the companies regarding the *continuous temporal customisation* of newsroom structures. All interviewees mention that the corporate newsroom starts from an initial (theory-based) prototype, which is then dynamically adapted to the company's communication needs. As explained by one expert, newsrooms are "learning organisation" or a "living process":

> The newsroom is not a fixed story, it is constantly evolving–like an ecosystem–and there is no such thing as a finished newsroom.

During the first one to two years, newsrooms focus on securing their position within the company. This involves active awareness raising—or "intra-company political campaigning", as one expert puts it—to accelerate the diffusion of the newsroom model as an organisational innovation. Several interviewees noted that during this time, the newsroom processes and structures are particularly "closely monitored" and undergo many changes until they are "safe enough to try". These changes are driven by an informal feedback culture and the experiences of key people in the newsroom. More formal evaluations of newsroom efficiency using management tools are largely omitted. Examples of changes at this stage include the creation or disbanding of new roles, shifting and clarifying responsibilities, and restructuring of the meeting protocols.

As the newsroom matures, its focus shifts outward, engaging more strongly with the larger corporate environment. This is best illustrated by the sharpening of the newsroom's topic orientation. On the one hand, new topics from the corporate environment flow into the newsroom through social listening activities. On the other hand, corporate newsrooms elaborate and expand their strategic topic management. In some cases, this consists of the systematic description and prioritisation of company topics (for example, as a differentiation between "umbrella" or "focus topics" or a more complex "topic architecture") along with the definition of corresponding routines based on communication priorities.

*Newsroom challenges and coping strategies*

Across interviews and companies, three common challenges and related coping strategies were repeatedly mentioned. The first and most critical challenge is to *establish and maintain a newsroom identity*. Inside the communication department, the transition to topic-oriented newsroom workflows with dynamic role-switching implies changes in the responsibilities of communication staff that not everyone is equally comfortable with or successful in adapting. As one interviewee noted, compared with traditional corporate communication models, newsrooms shed light on "blind spots of unproductivity". There is a need to foster a collective newsroom identity to adapt to the new ways of collaboration. This includes constant onboarding process and training, for instance, via internal workshops and targeted recruitment of new people. To preserve newsrooms' agility, companies actively promote an informal feedback culture and flatten hierarchies by attaching decision-making power to the technical expertise contained in specific roles.

Outside the communication department, many experts highlighted the necessity of a newsroom identity to tackle other departments' lack of involvement in newsroom processes. Such a dispassionate attitude is highly problematic as the topics that newsrooms tackle typically span the entirety of corporate operations. As showcased above, topic-oriented workflows are, by definition, predicated upon cross-departmental cooperation. In some cases, experts address this challenge by "polishing the door handles" of the various departments. Alternatively, companies can create the role of newsroom representatives in different departments, who serve as contact persons whom newsroom staff can approach. Similarly, in addition to the usual newsroom meetings, several companies have implemented sporadic "extended newsroom" gatherings, which are attended by external department members when needed.

A second challenge arises as the flipside of newsroom agility, which involves a *need for constant change management*. The flexibility with which roles within the newsroom are defined and redefined entails frequent changes in related workflows. Updating and refining the documentation of the various roles can prevent conflicts but is also time-consuming. Though communication experts do not offer a single solution to this problem, they highlight the value of frequent and informal feedback for finding the right balance between having the "courage to adapt" and avoiding the "energy drain" of uprooting routines.

Third, interviewees unanimously note the *comprehensive impact assessment* of communication as a central challenge for the newsroom model. Although the companies use many monitoring and analytical tools to measure communication impact, half of the interviewees wished to "close the feedback loop"—that is, to learn from responses to communicated content and integrate these learnings in the strategic decisions for future communication. Common outreach metrics are perceived as insufficient for understanding why certain stories are more read and liked. While the problem of impact measurement is not unique to the newsroom model, its strong emphasis on topic orientation and storytelling requires a more in-depth understanding of audience feedback to design a coherent communication strategy. Besides the compilation of cross-channel impact reports, often sourced by agencies, the interviewees expect to achieve a better understanding of communication impact by actively integrating data analysts in the newsrooms or by applying AI (see below for examples).

*Use cases of AI in corporate newsrooms*

When asked about the relevance of AI for newsroom discussions, interviewees' responses ranged from lukewarm ("The initial hype has died down already") to "pretty heated" with weekly discussions about AI applications. In terms of use cases, the interviews reveal a wide range of areas where AI is applied (or can be applied) in corporate newsrooms. Often, newsrooms see AI tools as "practical helpers" to support and enhance existing workflows. Here are the common examples of these *routine applications* that have become part of everyday corporate communications, depending on employees' affinity for technology:

- *Idea generation*: Several newsrooms use large language model (LLM)-based applications such as ChatGPT, Microsoft Copilot, or similar in-house solutions as creativity boosters. Typical use cases include brainstorming titles or ideas for stories.

- *Text production*: LLM-based applications and other forms of text-focused AI software (e.g. Deepl, Scribbr, or Grammarly) are used as automated "first-line tools" for routine proofreading and editing. Out-of-the-box tools sometimes must be enriched by integrating dialect- and subject-specific vocabularies that are particularly relevant to the multilingual and multi-dialect Swiss context.

- *Text translation*: Many newsrooms operate in multilingual environments and have long been using AI-enhanced translation applications (Deepl, in-house and specialised software with their own vocabulary) for targeting specific audiences in their native language.

- *Image/video editing*: Specialised AI applications are used for (audio-)visual content production (e.g. automated object recognition and editing in Adobe Photoshop or noise filters for reducing background sound in audio recordings)

- *Indexing*: LLM-based applications and other specialised tools used for content management are applied to extract keywords from (textual) data, such as earlier company publications, for indexing. Similarly, content on company websites is scanned and optimised for search engines using AI applications for search engine optimisation.

- *Impact measurement*: Many analytical tools for social media and media monitoring integrate AI-driven components (e.g. Argus, Sprinklr, Parse.Ly).

Several corporate newsrooms also conduct *AI pilot tests* to evaluate how "AI can be used for more radical use cases" that may restructure newsroom workflows. Below, we detail three such use cases that are of particular relevance for the newsroom model.

The first example is the use of AI as a *living corporate data archive*. As part of coordinating corporate communications, newsrooms are confronted with many internal and external information requests (sometimes in the four-digit range) every day. Answering these requests requires considerable resources, and many requests from customers, other departments, or journalists are similar and have already been addressed in the past. Three interviewees noted that they examine how AI can, as a first step, index answered requests, reports and other documents regarding corporate activities. In the second step, an LLM-based chatbot is deployed as a user-friendly access point to the knowledge archive. As one expert explains, such centralised corporate archives can amplify internal efficiency by facilitating the acquisition of information for content production:

> We are also working on indexing the image and sound material used. If I make a story about a specific place or person in the future, I will have direct access to all the material that already exists.

Another expert reports that she will no longer need assistance to scramble together corporate facts for publications or events; instead, "everything is simply pulled out of the AI corporate data universe". Although untested for now, interviewees also consider broadening access to the company's data archive for external inquiries, for instance, via an AI-operated service centre that can reduce human workload.

A second use case regards the *automated translation of stories for different communication channels.* Traditionally, topic managers in the newsrooms describe the story in the format of a specific channel (e.g. as a press release or a LinkedIn post). Instead of adapting stories for each channel manually, a story can be rewritten in a desired format via an LLM-based tool. One expert describes it like this:

> The expected benefit of AI should be massive in the medium term. At some point, you can say, for example: dear AI, here is the press release. Now make me some content, make me a tweet, make me a LinkedIn post.

Channel translation is also intended to help with the stylistic adaptation of stories. It can be beneficial if communicators are less familiar with the conventions and affordances of specific platforms such as TikTok or Instagram. Such a tool facilitates the topic-oriented logic of corporate newsrooms as they can relegate resources from channel management to producing high-quality content.

The third AI pilot taps into the potential of AI to facilitate *responsive impact measurement*. Three interviewees mentioned using LLM-based applications to aggregate impact reports as part of "structured experimentation". Such AI-facilitated aggregation can solve the problem of fragmented analytical newsroom tools, where each tool generates individual reports that must then be manually combined. One newsroom, for instance, tests the integration of the outputs of different tools via an in-house chatbot. Another newsroom uses predefined prompts for Perplexity AI, an LLM-based chatbot, to extract critical reactions to its publications from media coverage or user reviews and generate ideas for response strategies. The last example is an in-house scraping tool that tracks activities from relevant institutions (e.g. police reports) in real time and automatically forwards them to the newsrooms as push notifications via the internal communication platform (e.g. Slack).

*Regulation of AI: Risks and opportunities*

The growing use of AI within corporate newsrooms raises questions about how it can be organised and regulated. Interviewees note that currently, there is no stringent regulation of the use of AI in the newsrooms where they work. The interviews suggest that the companies are now transitioning between two phases: the initial AI discovery and test phase will be over soon, and a second phase of institutionalisation and AI urgently calls for regulation.

The initial AI discovery phase, where most Swiss newsrooms are currently located, focuses on exploring possible use cases, as described above. Impulses for the use of AI come mainly bottom-up from individual employees and the engagement with AI applications varies depending on motivation and affinity for technology within a specific company. Some newsrooms have created the roles of "super users" or "AI creators" to designate tech-savvy individuals who are actively involved in pilot studies and collaborate with the IT department. However, in general, the use of AI in this phase largely depends on the individual newsroom members.

After this initial build-up, the preliminary top-down institutionalisation of AI uses follows. Around half of the interviewees noted that they aim to develop an "AI framework" or "comprehensive AI guidelines" by the end of 2024. In addition, checklists are continuously being created for AI applications that are already used. For instance, one expert described the introduction of a decision grid for an AI-facilitated text translation tool. This grid defines which translation tasks AI can be used exclusively, partially, or not at all and prescribes the corresponding quality checks. These efforts show the first steps towards AI governance;

however, the comprehensive regulation of AI in corporate newsrooms seems difficult, given the speed of technical advancements and their adoption.

As for interviewees' opinions on the advantages and disadvantages of AI for the corporate newsroom, then, unsurprisingly, the main promise of AI relates to the potential efficiency gains. The professed aim is not the complete automation of work processes (e.g. to decrease the human workforce and cut the company's personnel). Instead, the purpose is to increase newsroom efficiency by optimising human work by integrating AI into communication processes in the background. There is also an expectation that AI can accelerate individual work routines that can "shorten the time from the initial impulse to the final communication output". It should amplify the creative potential of human communicators, as employees will have more time to explore new or more resource-demanding ideas. In several interviews, AI is explicitly connected to the opportunities for improving communication impact measurement and is perceived as a promising tool for facilitating social listening and aggregating impact indicators (see also the use case described above).

However, the extent to which AI tools can deliver the expected gains is currently uncertain, given the lack of experience and the rapidly changing state of the art of AI. Various interviews cite this uncertainty as a key risk. Currently, newsrooms are being "flooded with the AI boom", and it is difficult to anticipate "in which areas of AI will prove to be truly helpful and where it's just a gimmick". Missing interfaces and difficulties in integrating (proprietary) AI tools into the existing infrastructures also make investment decisions more difficult. On a human level, there are costs for (re)training employees and expanding AI-related skills. All interviews also cite data protection issues as a major risk. The lack of transparency about which data is collected when using (third-party) AI tools is critical for companies. A handful of the newsrooms have therefore started developing their own AI applications and implemented internal chatbots and translation software. Other newsrooms are currently planning this step.

The biggest concern, however, is related to quality assurance regarding the use of AI. One interviewee uses the example of community management to explain how AI could massively backfire: "we could badly burn our hands", if chatbots frustrate customers by failing to handle irony, not being able to navigate the complex dialectal landscape, inadequately conveying the company's tone. Experts emphasise the importance of manual checks and control frameworks to ensure quality for each use case (such as the grid for AI-assisted translations described above) and prevent the loss of control. There is a fundamental question of to what extent communication-related decisions should be entrusted to a "cold black box of AI" that has no clear sense of the company's best interests. Or, in the words of one expert:

> Communication can always have unintended consequences. I can trust my employees to do everything they can to prevent this scenario. That's not the case with AI.

## Discussion and Conclusion

In this study, we explored the features and processes of corporate newsrooms in Swiss companies across industry sectors and how corporate newsrooms tackle developments and challenges related to the rise of AI. Based on a set of semi-structured interviews with the experts, we found that newsrooms are understood as a mental and organisational model of communication. In contrast with textbook definitions (Moss, 2021), spatial anchoring is not a definitive criterion. Refining existing definitions (Seiffert-Brockmann *et al.*, 2021; Plesner and Raviola, 2016; Zerfass and Schramm, 2014), corporate newsrooms emerge as central control units for agile, topic-oriented, and role-based corporate communication that coordinate cross-departmental communication strategies and channels via regular meetings. In the logic of learning organisations (Senge, 2006), newsrooms are dynamic structures that are continuously calibrated in response to changing requirements and technologies.

We identified the construction of a collective newsroom identity and the balancing between optimising and stabilising the communication structures as two key challenges to the corporate newsroom management. The cultivation of informal feedback culture, interpersonal relationship management between departments, and capability-based redistribution of decision-making power can be seen as coping strategies based on informality (Koch and Denner, 2022); these strategies are lived examples of corporate newsrooms' recognition of "formal gaps to be filled by informal structures" (Hoffjann, 2024, p. 2024). The third challenge of including audience responses in communication impact measurement reflects a broader trend in corporate communication towards more rigorous social listening (Yavuz and Tire, 2023; Westermann and Forthmann, 2021). Innovative uses of AI can enhance the capabilities for integrating feedback from the public and help corporate newsrooms become not only learning but also listening organisations (Macnamara, 2020; Madsen and Andersen, 2023). As such, corporate newsrooms have the potential for co-creative integrated communication, as they not only streamline intra-organizational procedures but can incorporate consumer voices and market cultures (Johansen and Esmann Andersen, 2012), thereby strengthening their role and value within the company (Brockhaus and Zerfass, 2022)

In the context of innovation adoption, the interviews highlight that AI is already being used in corporate newsrooms for a wide range of tasks. Whereas most routine uses of AI involve skill- and rules-based tasks, more innovative pilot tests relate to the automation of knowledge-based tasks (Diakopoulos, 2019), such as the translation of content between communication channels or the transformation of company data into living archives.

However, besides the advantages of using AI, such as increasing efficiency, optimising processes and realising creative potential, the interviewees also highlighted multiple risks similar to those identified in journalistic newsrooms (Dralega, 2023; Munoriyarwa *et al.*, 2023): uncertain returns for investments, data protection problems and difficulties with quality assurance. These observations highlight an acute need for governance and a nuanced understanding of AI-driven communication tools to realise their potential and minimise possible negative effects.

It is also important to mention several limitations of the current research and possible directions for future studies. The first of them regards the small sample of interviewees that has implications for the generalizability of the findings for the Swiss corporate environment. It also impedes the meaningful comparison of the differences in the implementation of newsrooms and the attitudes towards AI across different corporate sectors; however, such a comparison constitutes an important future direction for corporate communication research. Another limitation regards the focus on the newsroom managers who are responsible for leading the newsrooms, whereas the perspectives of other members of the newsroom staff remain unaccounted for. Finally, in this study, we focused on a single national case, whereas insights into other regional and cultural contexts in which the newsroom model may be adopted are needed. Future research will benefit from expanding the selection of case studies, especially by looking beyond the Global North and conducting more comparative research on different aspects of newsrooms' adoption and functioning.

**Funding statement**

The project has been financially supported by the Newsroom Communication AG. The funder provided non-binding feedback for the study design and facilitated data collection by expediting selection and contacting interviewees for the study. Besides this, the funder had no role in data collection and analysis, the decision to publish, or manuscript preparation.

**References**

Al-Warith, G. and Moss, C. (2021), "Case study Dutch national Police: building trust with a corporate newsroom", in Moss, C. (Ed.), *The corporate newsroom: steering companies efficiently through communication*, Springer, Wiesbaden, pp. 129-136.

Bastian, M., Helberger, N., and Makhortykh, M. (2021), "Safeguarding the journalistic DNA: Attitudes towards the role of professional values in algorithmic news recommender designs", *Digital Journalism*, Vol. 9 No. 6, pp. 835-863. doi: 10.1080/21670811.2021.1912622

Bento, F., Tagliabue, M. and Lorenzo, F. (2020), "Organizational silos: a scoping review informed by a behavioral perspective on systems and networks", *Societies*, Vol. 10 No. 56, pp. 1-27. doi: 10.3390/soc10030056.

Brockhaus, J. and Zerfass, A. (2022), "Strengthening the role of communication departments: a framework for positioning communication departments at the top of and throughout organizations", Corporate Communications: An International Journal, Vol. 27 No. 1, pp. 53-70. doi: 10.1108/CCIJ-02-2021-0021

Brockhaus, J., Buhmann, A. and Zerfass, A. (2022), "Digitalization in corporate communications: understanding the emergence and consequences of CommTech and digital infrastructure", Corporate Communications: An International Journal, Vol. 28 No. 2, pp. 274-292. doi: 10.1108/CCIJ-03-2022-0035.

Buggisch, C. (2021), "Case Study DATEV: the introduction of a corporate newsroom as a change project", in Moss, C. (Ed.), *The corporate newsroom: steering companies efficiently through communication*, Springer, Wiesbaden, pp. 117-127.

Capriotti, P., Zeler, I. and Camilleri, M.A. (2021), "Corporate communication through social networks: the identification of the key dimensions for dialogic communication", in Camilleri, M.A. (Ed.), *Strategic Corporate Communication in the Digital Age*, Emerald Publishing, Bingley, UK, pp. 33-51.

Coombs, W. T. and Holladay, J. S. (2012), "The paracrisis: the challenges created by publicly managing crisis prevention", *Public Relations Review*, Vol. 38 No. 3, pp. 408-415. doi: 10.1016/j.pubrev.2012.04.004.

Coombs, T. and Holladay, S. (2023), "Digital corporate communication and paracrises and AI", in Luoma-aho, V. and Badham, M (Ed.s.), *Handbook on Digital Corporate Communication*, Edward Elgar Publishing, Cheltenham, UK, pp. 165-178. doi: 10.4337/9781802201963.00022.

Cornelissen, J. P. (2020), *Corporate communication*, Sage, Thousand Oaks, CA.

Diakopoulos, N. (2019), *Automating the news: how algorithms are rewriting the media*, Harvard University Press, Cambridge, MA.

de Regt, A., Plangger, K. and Barnes, S. J. (2021), "Virtual reality marketing and customer advocacy: transforming experiences from story-telling to story-doing", *Journal of Business Research*, Vol. 136, pp. 513-522. doi: 10.1016/j.jbusres.2021.08.004.

Dolev, E. L., Lutz, C. F. and Aepli, N. (2024), "Does Whisper understand Swiss German? An automatic, qualitative, and human evaluation", paper presented at VarDial 2024 (the eleventh Workshop on NLP for Similar Languages, Varieties and Dialects), 20.6.2024, Mexico City, available at: https://doi.org/10.48550/arXiv.2404.19310.

Dou, W., Lim, K. H., Su, C., Zhou, N. and Cui, N. (2010), "Brand positioning strategy using search engine marketing", *MIS Quarterly*, Vol. 34, No. 2 pp. 261-279. doi: https://doi.org/10.2307/20721427.

Dralega, C. A. (2023), "AI and the algorithmic-turn in journalism practice in Eastern Africa: perceptions, practice and challenges", in Dralega, C. (Ed.), *Digitisation, AI and algorithms in African journalism and media contexts*, Bingley, UK, pp. 33-52.

Hoffjann, O. (2024), "Informality in strategic communication: making the case for a forgotten concept", *Corporate Communications: An International Journal*, Vol. 29 No. 2, pp. 206-220. doi: https://doi.org/10.1108/CCIJ-03-2023-0028.

Holzinger, T. and Sturmer, M. (2012), *Im Netz der Nachricht: die Newsroom-Strategie als PR-Roman*, Springer, Berlin and Heidelberg.

Johansen, S. T. and Andersen, S. E. (2012), "Co-creating ONE: rethinking integration within communication", *Corporate Communications: An International Journal*, Vol. 17 No. 3, pp. 272-288. doi: 10.1108/13563281211253520

Keel, G. and Niederhäuser, M. (2016). *Corporate Newsrooms in der Schweiz. Ergebnisse einer Befragung von Schweizer Unternehmen und Verwaltungen,* study report by IAM Institute of Applied Media Studies, Zürich University of Applied Sciences (ZHAW).

Koch, T. and Denner, N. (2022), "Informal communication in organizations: work time wasted at the water-cooler or crucial exchange among co-workers?", *Corporate Communications: An International Journal*, Vol. 27 No. 3, pp. 494-508. doi: 10.1108/CCIJ-08-2021-0087

Madsen, V. T. and Andersen, H. T. (2024), "A move to the bright side? When journalism is invited into internal communication", *Corporate Communications: An International Journal*, Vol. 29 No. 2, pp. 221-237. doi: 10.1108/CCIJ-12-2022-0156

Martinelli, R. (2019), "Corporate mobile communication: challenges and reflections in an environment of connected employees", in Thornton, G.S., Mansi, V.R., Carramenha, B. and Cappellano T. (Ed.s.), *Strategic employee communication: building a culture of engagement*, Springer, Cham, pp. 283-294.

Moss, C., Westermann, A. and Ghorbani, M. (2019). *Der Corporate Newsroom beginnt im Kopf*, study report by BRMI@ISM and Mediamoss.

Moss, C. (2021, Ed.), *The corporate newsroom: steering companies efficiently through communication*, Springe, Wiesbaden.

Moss *et al*., 2019; Keel and Niederhäuser, 2016

Moss, C. (2021), "Case Study R+ V Insurance: meeting new challenges with a corporate newsroom", in Moss, C. (Ed.), *The corporate newsroom: steering companies efficiently through communication*, Springer, Wiesbaden, pp. 137-143.

Munoriyarwa, A., Chiumbu, S. and Motsaathebe, G. (2021), "Artificial intelligence practices in everyday news production: the case of South Africa's mainstream newsrooms", *Journalism Practice*, Vol. 17 No. 7, pp.1374-1392. https://doi.org/10.1080/17512786.2021.1984976.

Nyagadza, B. (2022), "Search engine marketing and social media marketing predictive trends", *Journal of Digital Media and Policy*, Vol. 13 No. 3, pp. 407-425. doi: 10.1386/jdmp_00036_1

Plesner, U. and Raviola, E. (2016), "Digital technologies and a changing profession: new management devices, practices and power relations in news work", *Journal of Organizational Change Management*, Vol. 29 No. 7, pp. 1044-1065. doi: 10.1108/JOCM-09-2015-0159

Radford, A., Kim, J. W., Xu, T., Brockman, G., McLeavey, C. and Sutskever, I. (2023), "Robust speech recognition via large-scale weak supervision", in Krause, A., Brunskill, E., Kyunghyun, C., Engelhardt, B., Sabato S. and Scarlett, J. (Ed.s) *International conference on machine learning, PMLR*, Vol. 202, pp.28492-28518.

Ragas, M. and Ragas, T. (2021), "Understanding agile for strategic communicators: foundations, implementations, and implications", *International Journal of Strategic Communication*, Vol. 15 No. 2, pp. 80-92. doi: 10.1080/1553118X.2021.1898147

Reyes-de-Cózar, S., Pérez-Escolar, M. and Navazo-Ostúa, P. (2022), "Digital Competencies for New Journalistic Work in Media Outlets: A Systematic Review", *Media and Communication*, Vol. 10, No. 1, pp. 27-42. doi: 10.17645/mac.v10i1.4439

Seiffert-Brockmann, J., Einwiller, S., Ninova-Solovykh, N. and Wolfgruber, D. (2021), "Agile content management: strategic communication in corporate newsrooms", *International Journal of Strategic Communication*, Vol. 15 No. 2, pp. 126-143. doi: 10.1080/1553118X.2021.1910270

Senge, P. M. (2006), *The fifth discipline: the art and practice of the learning organization*, Broadway Business, New York, NY.

Singh, S. and Sonnenburg, S. (2012), "Brand performances in social media", *Journal of Interactive Marketing*, Vol. 26 No. 4, pp.189-197. doi: 10.1016/j.intmar.2012.04.00

Spanos, M. (2021), "Brand storytelling in the age of artificial intelligence", *Journal of Brand Strategy*, Vol. 10 No. 1, pp.6-13.

Sun, Y., Chen, J. and Sundar, S. S. (2024), "Chatbot ads with a human touch: a test of anthropomorphism, interactivity, and narrativity", *Journal of Business Research*, Vol. 172, No. 114403, pp. 1-13. doi: 10.1016/j.jbusres.2023.114403

Vernuccio, M. (2014), "Communicating corporate brands through social media: an exploratory study", *International Journal of Business Communication*, Vol. 51 No. 3, pp. 211-233. doi: 10.1177/2329488414525400

Vogler, D. and Badham, M (2023), "Digital corporate communicatin and media relations", in Luoma-aho, V. and Badham, M (Ed.s.), *Handbook on Digital Corporate Communication*, Edward Elgar Publishing, Cheltenham, UK, pp. 51–63. doi: 10.4337/9781802201963.00013

Westermann, A. and Forthmann, J. (2021), "Social listening: a potential game changer in reputation management How big data analysis can contribute to understanding stakeholders' views on organisations", *Corporate Communications: An International Journal*, Vol. 26 No. 1, pp. 2-22. doi: 10.1108/CCIJ-01-2020-0028

Westerlund, M. (2019), "The emergence of deepfake technology: a review", *Technology Innovation Management Review*, Vol. 9 No. 11, pp. 39-52.

Xu, M., David, J. M. and Kim, S. H. (2018), "The fourth industrial revolution: opportunities and challenges", *International Journal of Financial Research*, Vol. 9 No. 2, pp. 90-95. doi: 10.5430/ijfr.v9n2p90

Yavuz, Ş. and Tire, E. (2023), "A survey of corporate communication professionals' perspective on social listening and analytics", *Corporate Communications: An International Journal*, Vol. 28 No. 4, pp. 564-581. doi: 10.1108/CCIJ-03-2022-0036

Zerfass, A. and Schramm, D. M. (2014), "Social media newsrooms in public relations: a conceptual framework and corporate practices in three countries", *Public Relations Review*, Vol. 40 No. 1, pp. 79-91. doi: 10.1016/j.pubrev.2013.12.003

Zerfass, A., Hagelstein, J. and Tench, R. (2020), "Artificial intelligence in communication management: a cross-national study on adoption and knowledge, impact, challenges and risks", *Journal of Communication Management*, Vol. 24 No. 4, pp. 377-389. doi: 10.1108/JCOM-10-2019-0137

Zerfass, A. and Link, J. (2023), "Business models for communication departments: a comprehensive approach to analyzing, explaining and innovating communication management in organizations", *Journal of Communication Management*, Ahead of print. doi: 10.1108/JCOM-02-2023-0027

**Appendix: Interview guide**

**1.Block: Briefing & icebreaker**
Demographic questions [either during the interview or in the e-mail]: age and gender [often requested by the journal reviewers these days in the aggregate anonymized form]

Brief: Note that we are never interested in any right or wrong answers. We'd like to gather ideas from practitioners/experts like yourself to get a diverse and honest insight. Your answers will be anonymized so that they cannot be connected to your name or [company X]. We will not "report back" to your company in any way about the specific content of this interview. We will audiorecord and transcribe this interview to store the anonymized transcript on a secure server. Is this okay with you?

After consent: Great, I will start the recording now.

- Q1 [icebreaker]: To start off, can you tell us a bit about your current position? How long are you working in [company X] and what are your main tasks?

**2. Block: Characteristics of Corporate Newsroom**
As we mentioned in the email, we are particularly interested in your experience of working with a corporate newsroom model.
- Q2.1: First we'd like to get a general overview of the newsroom at [company X]:
  - How many people work in the newsroom structure?
  - When did you implement a newsroom structure?
  - What was the motivation for implementing it?
  - How did you go about implementing it?
- Q2.2: Can you describe what the main tasks and purposes of the newsroom are?
  - Follow-up: tell us a bit more about what actually "happens" in your newsroom?
  - Follow-up: What are the different roles in the newsroom? What the different roles and tasks are?

- Q2.3: If you take a step back-How would you describe your experience with the newsroom so far?
  - Follow-up: what are the main benefits of having a newsroom structure in your company?
  - Follow-up: And what are downsides?
  - Follow-up: Are there things that have not worked out as planned?
  - Follow-up: Are there aspects of the newsroom model that you plan on implementing or improving in the future?
  - Follow-up: What is currently the "hot topic" in your newsroom structure (e.g., resources, data analysis, AI)

- Q2.4: What criteria would you (or the company) use to evaluate the success of a newsroom model at [company X]?
  - Who does the evaluation (if at all)?
  - And what tools are used for this purpose?

- o   What do you think of this evaluation?
- o   Is there something these criteria might not capture or see?

## 3. Block: Use of AI

Today, many companies increasingly integrate artificial intelligence (AI) in their communication strategies and tools.

- Q3.1: How does this integration of AI tools look like in [company X] in general?
  - o   (if unmentioned): can you give examples of specific AI tools that are in use?
- Q3.2: How about AI in the newsroom: Are there any AI tools in the newsroom?
  - o    can you give examples of such tools/software and what they are used for?
- Q3.3: What benefits and dangers do you think these AI tools bring for the newsroom? (follow-up: and for company X in general?)
- Q3.4: In your view, is there anything "specifically Swiss" about how newsroom models are adopted in Swiss companies?
  - o   (if unmentioned/unclear): Does the Swiss market make it particularly easy/difficult to implement such a model?
  - o   (if hesitation/unmentioned): do you think similar companies in other countries are doing things differently than here?

## 4. Block: Newsroom values

The concept of a newsroom originates from traditional journalism and news organisations.

- Q4.1: In your personal view, what are the differences between a corporate and a journalistic newsroom?
  - o   (if unmentioned): what might be similar to traditional journalistic newsrooms?
  - o   Q4.2: Now we'd like to change things up a bit and play a small "game". In the past, we conducted similar interviews with journalists and asked them about the importance of professional values when it comes to AI in their work. Now, we would like to do a similar exercise with representatives of corporate newsrooms. Imagine that we can select three professional values from this pool of cards [show cards or a slide] which AI tools will support and enhance. Which values will you choose [we do not define these values, because otherwise we can influence your views, so just select them based on how you understand them]?

Values list: 1) providing a public service to people as citizens and/or consumers; 2) watchdog of political and other elites; 3) legitimacy; 4) commitment to truth; 5) objectivity; 6) fairness; 7) professional distance/detachment; 8) impartiality; 9) neutrality; 10) credibility; 11) diversity; 12) cultural plurality; 13) independence; 14) (editorial) autonomy; 15) actuality; 16) speed; 17) depth; 18) inclusiveness; 19) transparency; 20) accountability; 21) social responsibility; 22) control; 23) responsiveness; 24) social cohesion;

- Q4.3: Now, let us have a look at the three values you selected. Can you tell very briefly about how you understand each of them and why you find it important?
  - o   (Follow-up if unclear/incomplete): how do you think AI helps supporting these values?

Q5: Before we end this conversation: what is the most important thing that the scientific community needs to know about corporate newsrooms?